\documentclass[aps,prl,preprint,groupedaddress]{revtex4-1}

\usepackage{theorem,amsmath,amssymb}

\newcommand{\R}{{\mathbb R}}
\newcommand{\eps}{\varepsilon}
\newcommand{\G}{{\mathcal G}}

\newcommand{\cA}{{\mathord{\cal A}}}
\newcommand{\cH}{{\mathord{\cal H}}}

\newtheorem{theorem}{Theorem}%[chapter]
\theoremstyle{theorem}
\newtheorem{proposition}[theorem]{Proposition}

\newtheorem{corollary}[theorem]{Corollary}
\theoremstyle{remark}

\begin{document}

%\preprint{}

\title{Kohn-Sham Theory in the Presence of Magnetic Field}


\author{Andre Laestadius}
\email{andrela@math.kth.se}
\thanks{Department of Mathematics, KTH Royal Institute of Technology, Sweden}
\affiliation{Department of Mathematics, KTH Royal Institute of Technology, Sweden}


\date{\today}

\begin{abstract}
In the well-known Kohn-Sham theory in Density Functional Theory, a fictitious non-interacting system 
is introduced that has the same particle density as a system of $N$ electrons subjected to mutual 
Coulomb repulsion and an external electric field. For a long time, the treatment of the kinetic energy was not correct and the theory 
was not well-defined 
for $N$-representable particle densities. In the work of [Hadjisavvas and Theophilou, Phys. Rev. A, 1984, 30, 2183], a rigorous 
Kohn-Sham theory for $N$-representable particle densities was 
developed using the Levy-Lieb functional. Since a Levy-Lieb-type functional can be defined for Current Density Functional Theory formulated 
with the paramagnetic current density, we here develop a rigorous $N$-representable Kohn-Sham approach for interacting 
electrons in magnetic field. 
Furthermore, in the one-electron case, criteria for $N$-representable particle densities to be 
$v$-representable are given.
\end{abstract}

%\keywords{}

\maketitle

\section{I. INTRODUCTION}
In the fundamental paper by Hohenberg and Kohn \cite{HK64}, the theoretical foundation of Density Functional Theory (DFT) was established. 
The Hohenberg-Kohn theorem states that, for a quantum mechanical system, the particle density $\rho$ determines the scalar potential 
of the system up to a constant. 
Subsequently, Kohn and Sham provided an algorithm \cite{KS65}, the so-called Kohn-Sham equations, for computing the density. These equations bear much 
resemblance to the Hartree-Fock integro-differential equations. % in that they all have to be solved iteratively until self-consistency. 
The idea of Kohn and Sham was to introduce a fictitious system of non-interacting particles that has the same particle density as the real 
interacting system. This is achieved by means of the exchange-correlation functional, which accounts for the non-classical 
two-particle interactions and the residual between the interacting and non-interacting kinetic energy. However, 
this functional remains unknown.

In the work of Hadjisavvas and Theophilou \cite{Hadjisavvas84}, a mathematically rigorous Kohn-Sham approach was developed. 
The importance of this work relies on the fact that $N$-representability can be guaranteed for a proper wavefunction, whereas $v$-representability 
cannot. This means, in principle, that any $v$-representable formalism is unjustified.

In the presence of a magnetic field, no Hohenberg-Kohn theorem exists at the present time (that is valid for any number of 
electrons). For the formulation of 
Current Density Functional Theory (CDFT) that uses the paramagnetic current density $j^p$, it is well-known that the density pair 
$(\rho,j^p)$ does not determine the scalar potential and vector potential of the system \cite{Vignale2002}. Counterexamples have been 
constructed that show that a ground-state can come from two different Hamiltonians \cite{AndreMichael,Vignale2002}. Thus, the particle density 
$\rho$ and the paramagnetic current density $j^p$ do not fully determine the Hamiltonian. For a many-electron system, 
neither proof nor counterexample exists 
so far in the literature for a Hohenberg-Kohn theorem formulated with the total current density $j$ \cite{AndreMichael,Tellgren}. 
In the one-electron case, on the other hand, it is possible to give a direct proof that $\rho$ and $j$ determine the scalar and vector potential up to a 
gauge transformation \cite{AndreMichael,Tellgren}.

However, since the density pair $(\rho,j^p)$ determines the (possibly degenerate) ground-state(s) of the system \cite{Vignale87,AndreMichael},
this work aims at continue the $N$-representable approach of \cite{Hadjisavvas84} and develop 
a rigorous Kohn-Sham approach for CDFT formulated with the paramagnetic current density $j^p$.

\section{II. CURRENT DENSITY FUNCTIONAL THEORY}
We will in this paper consider a system of $N$ interacting electrons subjected to both an electric and a 
magnetic field. The system's Hamiltonian is given by (in suitable units)
\begin{align*}
H(v,A)=\sum_{k=1}^N \left((i\nabla_k - A(x_k))^2 + v(x_k) \right) + \sum_{1\leq k<l\leq N} |x_k - x_l|^{-1},
%\label{Hamil}
\end{align*}
where $v(x)$ is the scalar potential and $A(x)$ the vector potential. The magnetic field is computed from 
$B(x) = \nabla \times A(x)$.
%We require that $v\in L^{3/2}(\R^3) + L^\infty(\R^3)$ and $A^k\in L^\infty(\R^3)$, where $A^k$ is the $k$:th component of $A$.
Throughout we will assume that the ground-state is non-degenerate, i.e., $\text{dim ker}(e_0-H(v,A))= 1$, where 
$e_0$ is the lowest eigenvalue of $H(v,A)$. %, and (ii) the ground-state $\psi_0$ of $H(v,A)$ has the weak unique continuation property (WUCP), 
%that is $\psi_0\neq 0$ almost everywhere (a.e.).

%Throughout the following will be assumed: (i) the ground-state is non-degenerate, i.e., $\text{dim ker}(e_0-H(v,A))= 1$, where 
%$e_0$ is the lowest eigenvalue of $H(v,A)$, and (ii) the ground-state $\psi_0$ of $H(v,A)$ has the weak unique continuation property (WUCP), 
%that is $\psi_0\neq 0$ almost everywhere (a.e.).% and (iii) the magnetic field $B=\nabla\times A$ has compact support and $A^k$ belongs to $L^\infty(\R^3)$.

\subsection{A. Preliminaries}
To begin with, some mathematical concepts needed for the forthcoming discussion are introduced. We first mention some relevant function spaces. 
If for some $p\in [1,\infty)$ a function $f$ satisfies $\int_{\R^n} |f|^p<\infty$, then $f$ belongs to the normed space $L^p(\R^n)$ with 
norm $||f||_{L^p(\R^n)} = (\int_{\R^n} |f|^p)^{1/p}$. In the case $p=\infty$, we say $f\in L^\infty(\R^n)$ if 
\[||f||_{L^\infty(\R^n)} = \text{ess} \sup \{|f|\, |x\in \R^n \}<\infty.\] 
Furthermore, $f\in L^2(\R^n)$ is said to belong to the Hilbert space $\cH^1(\R^n)$ if 
\[||f||_{\cH^1(\R^n)}^2 = \int_{\R^n}|f|^2 + \int_{\R^n}|\nabla f|^2<\infty.\] 
Let $B_R = \{x\in\R^n| \,|x|\leq R \}$ for $R>0$. Then 
$f\in L_{\text{loc}}^1(\R^n)$ whenever $\int_{B_R} |f|<\infty$ for any $B_R$. For a vector $u$ such that 
$(u)_l\in L^p$, $l=1,2,3$, we write $u\in (L^p)^3$.

We say that a sequence $\{\psi_k\}\subset L^p(\R^n)$ converges 
in $L^p(\R^n)$-norm to $\psi \in L^p(\R^n)$ if $\int_{\R^n} |\psi_k - \psi|^p \rightarrow 0$ as $k\rightarrow \infty$, and we write 
$\psi_k\rightarrow \psi$. For the Hilbert space $L^2(\R^n)$, with inner product $(\psi,\phi)_{L^2(\R^n)} = \int_{\R^n}\psi^*\phi$, 
we say that $\{\psi_k\}\subset L^2(\R^n)$ converges weakly 
to $\psi \in L^2(\R^n)$ if $(\psi_k,\phi)_{L^2(\R^n)}  \rightarrow (\psi,\phi)_{L^2(\R^n)}$ as $k\rightarrow \infty$ 
for all $\phi \in L^2(\R^n)$, and we write $\psi_k\rightharpoonup \psi$. 
For weak convergence in $\cH^1(\R^n)$, we require $(\psi_k,\phi)_{\cH^1(\R^n)} \rightarrow (\psi,\phi)_{\cH^1(\R^n)}$ as $k\rightarrow \infty$ for all $\phi\in \cH^1(\R^n)$, 
where the inner product of $\cH^1(\R^n)$ is given by 
$(\psi,\phi)_{\cH^1(\R^n)} = \int_{\R^n }\psi^* \phi +  \int_{\R^n }\nabla \psi^*\cdot \nabla \phi$. Weak convergence 
on $\cH^1(\R^n)$ implies weak convergence in the $L^2(\R^n)$ sense. A functional $f$ is said to be weakly lower semi continuous if 
$\psi_k\rightharpoonup \psi$ implies $\liminf_{k\rightarrow \infty} f(\psi_k)\geq f(\psi)$. In particular, 
$\liminf_{k\rightarrow\infty} || \psi_k ||_{L^2(\R^n)}\geq|| \psi ||_{L^2(\R^n)}$ if $\psi_k\rightharpoonup \psi$ weakly in $L^2(\R^n)$.\\

For a fixed particle number $N$, define the set of proper wavefunctions to be
\begin{align*}
W_N = \{\psi\in \cH^1(\R^{3N})|\text{$\psi$ antisymmetric and $||\psi||_{L^2(\R^{3N})}=1$}\}
\end{align*}
and let the ground-state energy of $H(v,A)$ be given by
\begin{align*}
e_0(v,A) = \inf\{ \mathcal{E}_{v,A}(\psi)|\psi\in W_N \},
\end{align*}
where
\begin{align*}
\mathcal{E}_{v,A}(\psi) = \sum_{k=1}^N \left(\int_{\R^{3N}} |(i\nabla_k - A(x_k))\psi|^2 + \int_{\R^{3N}} |\psi|^2v(x_k) \right)
+\sum_{1\leq k<l\leq N} \int_{\R^{3N}} |\psi|^2|x_k - x_l|^{-1}.
\end{align*}
We will define the inner-product $(\psi,H(v,A)\psi)_{L^2}$ as the number $\mathcal{E}_{v,A}(\psi)$ for $\psi\in W_N$, 
even if $H(v,A)\psi\notin L^2$.

The particle and paramagnetic current density for $\psi\in W_N$ are computed from
\begin{align*}
\rho_\psi(x) &=N \int_{\R^{3(N-1)}} |\psi(x,x_2,\dots,x_N)|^2 dx_2\dots dx_N,\nonumber \\
j_\psi^p(x) &=N \,\text{Im} \int_{\R^{3(N-1)}} \psi^*(x,x_2,\dots,x_N)\nabla_x \psi (x,x_2,\dots,x_N)dx_2\dots dx_N,
\end{align*}
respectively. We will use the notation $\psi\mapsto (\rho,j^p)$ to mean $\rho_\psi=\rho$ and $j_\psi^p=j^p$. Furthermore, we 
shall use the notation $H_0$ for the Hamiltonian $H(v,A)$ when the potential terms are set to zero, i.e.,
\begin{align*}
(\psi,H_0\psi)_{L^2} = \sum_{k=1}^N \int_{\R^{3N}}|\nabla_k\psi|^2 + \sum_{1\leq k<l\leq N} \int_{\R^{3N}} |\psi|^2|x_k - x_l|^{-1}.
\end{align*}
Note that
\begin{align*}
\mathcal{E}_{v,A}(\psi) = (\psi,H(v,A)\psi)_{L^2} = (\psi,H_0\psi)_{L^2} + 2\int_{\R^3}j_\psi^p\cdot A + \int_{\R^3} \rho_\psi (v + |A|^2),
\end{align*}
which follows from a direct computation.

\subsection{B. $N$-representable DFT}
A $v$-representable particle density is a density $\rho$ that satisfies $\rho= \rho_\psi$ and where $\psi$ is the ground-state 
of some $H(v)$. (We will use the notation $H(v)=H(v,0)$ and $e_0(v) = e_0(v,0)$ when not considering magnetic fields.) 
The set of $N$-representable particle densities is given by \cite{Lieb83}
\[I_N=\Big\{ \rho| \rho\geq 0 ,\int_{\R^3}\rho =N,\rho^{1/2}\in \cH^1(\R^3)\Big\}. \]
As demonstrated by Englisch and Englisch in \cite{Englisch}, 
not every $N$-representable particle density is $v$-representable.
For $\rho\in I_N$, the Levy-Lieb functional 
\begin{align*}
F_{LL}(\rho) =\inf\{(\psi,H_0\psi)_{L^2}| \psi\in W_N,\psi \mapsto \rho \}
\end{align*}
is well-defined. As was proven in \cite{Lieb83} (Theorem 3.3), there exists a $\psi_0\in W_N$ such that $F_{LL}(\rho) =(\psi_0,H_0\psi_0)_{L^2}$ 
and $\rho_{\psi_0} = \rho$. The functional $F_{LL}(\rho)$ extends the Hohenberg-Kohn functional to $N$-representable densities, and 
for the ground-state energy we have
\begin{align*}
e_0(v) =\inf\Big\{F_{LL}(\rho) + \int_{\R^3} \rho v| \rho \in I_N \Big\}.
\end{align*}
Note that the number $e_0(v)$ is well-defined for $v\in L^{3/2}(\R^3)  + L^\infty(\R^3)$ even if $H(v)$ does not have a ground-state. 
($\int_{\R^3}\rho v$ is finite for all $\rho\in I_N$, since $I_N\subset L^1(\R^3) \cap L^3(\R^3)$, see \cite{Lieb83}.)

\subsection{C. $N$-representable CDFT}
A density pair $(\rho,j^p)$ is said to be $v$-representable if there exists a $\psi$ that is the ground-state of some Hamiltonian $H(v,A)$ 
such that $\rho = \rho_\psi$ and $j^p=j_\psi^p$. We denote this set of densities $\cA_N$, i.e.,
\begin{align*}
\cA_N = \{ (\rho,j^p)| \text{there exists a $H(v,A)$ with ground-state $\psi$ such that $\psi\mapsto (\rho,j^p)$}\}.
\end{align*}
Now, assume that $H(v_1,A_1)$ and $H(v_2,A_2)$ have the ground-states $\psi$ and $\phi$, respectively. Then from Theorem 9 in \cite{AndreMichael}, 
if $\psi\mapsto (\rho,j^p)$ and $\phi\mapsto (\rho,j^p)$, it follows that $\psi = \text{const.}\,\phi$. For $(\rho,j^p)\in \cA_N$, 
let $\psi_{\rho,j^p}$ denote the ground-state of some $H(v,A)$ such that $\psi\mapsto (\rho,j^p)$. Then the generalized 
Hohenberg-Kohn functional 
\begin{align*}
F_{HK} (\rho,j^p) = (\psi_{\rho,j^p}, H_0 \psi_{\rho,j^p})_{L^2}
\end{align*}
is well-defined on $\cA_N$. Furthermore (Theorem 2 in \cite{Andre}), 
\begin{align*}
e_0(v,A) = \min \Big\{ F_{HK} (\rho,j^p) + 2\int_{\R^3}j^p\cdot A + \int_{\R^3} \rho (v + |A|^2) \Big| (\rho,j^p)\in \cA_N\Big\}
\end{align*}
for $(v,A)\in V_N$, where
\begin{align*}
V_N = \{(v,A)| \text{$H(v,A)$ has a unique ground-state} \}.
\end{align*}
However, a $\psi\in W_N$ may be such that $(\rho_\psi,j_\psi^p)\notin \cA_N$. From Proposition 3 in \cite{Andre}, 
$\psi\in W_N$ implies that $\psi\mapsto (\rho,j^p)\in Y_N$, where
\begin{align*}
Y_N = \Big\{ (\rho,j^p)| \rho\geq 0 ,\int_{\R^3}\rho =N,\rho^{1/2}\in \cH^1(\R^3), j^p\in (L^1(\R^3))^3,
\int_{\R^3}|j^p|^2\rho^{-1}<\infty  \Big\}.
\end{align*}
The set $Y_N$ is referred to as the set of $N$-representable density pairs $(\rho,j^p)$. It is a convex set and $\cA_N \subsetneq Y_N$ 
(Proposition 4 in \cite{Andre}). For $(\rho,j^p)\in Y_N$, define as in \cite{Andre}
\begin{align*}
Q(\rho,j^p) =\inf\{(\psi,H_0\psi)_{L^2}| \psi\in W_N,\psi \mapsto (\rho,j^p) \}.
\end{align*}
The functional $Q(\rho,j^p)$ is the generalization of the Levy-Lieb functional $F_{LL}(\rho)$. It also depends on the paramagnetic 
current density $j^p$. The functional $Q(\rho,j^p)$ inherits many properties of $F_{LL}(\rho)$: by Theorem 5 and Theorem 6 
in \cite{Andre}, we have (i) $Q(\rho,j^p) = F_{HK}(\rho,j^p)$ for $(\rho,j^p)\in \cA_N$, (ii) 
there exists a $\psi_m\in W_N$ such that $Q(\rho,j^p) = (\psi_m,H_0\psi_m)_{L^2}$ and where $\psi_m\mapsto (\rho,j^p)$, and (iii) 
\begin{align*}
e_0(v,A) = \inf \Big\{ Q (\rho,j^p) + 2\int_{\R^3}j^p\cdot A + \int_{\R^3} \rho (v + |A|^2) \Big| (\rho,j^p)\in Y_N\Big\}.
\end{align*}

In \cite{Hadjisavvas84}, $F_{LL}(\rho)$ was used to obtain a rigorous Kohn-Sham theory for $N$-representable densities. Before generalizing this to 
CDFT formulated with $j^p$, we shall discuss the following question raised in \cite{Hadjisavvas84}: since a $\psi_0\in W_N$ exists 
such that $F_{LL}(\rho) = (\psi_0,H_0\psi_0)_{L^2}$ and $\psi_0\mapsto \rho$, does $\psi_0$ satisfy any Schr\"odinger equation, i.e., 
is there a $v(x)$ such that $H(v)\psi=e\psi$?

\section{III. CHARACTERIZATION OF $V$-REPRESENTABLE PARTICLE DENSITIES}
We start be stating the mentioned result of Lieb (Theorem 3.3 in \cite{Lieb83}) for the functional $F_{LL}(\rho)$.
\begin{theorem}
There exists a $\psi_0$ in $W_N$ such that 
for $\rho\in I_N$, $F_{LL}(\rho)=(\psi_0,H_0\psi_0)_{L^2}$ and $\rho_{\psi_0}=\rho$.
\label{LiebThm1}
\end{theorem}
Let $\rho\in I_N$. In light of Theorem \ref{LiebThm1}, if the minimizer $\psi_0$ would be the ground-state of some Hamiltonian $H(v)$, 
then $\rho$ would be $v$-representable. However, since the $v$-representable densities are a proper subset of the $N$-representable ones \cite{Englisch}, 
there exists $\rho\in I_N$ such that the corresponding minimizer $\psi_0$ is not the ground-state of any Hamiltonian $H(v)$. Also note that, if 
$\rho$ is $v$-representable, then the minimizer $\psi_0$ is also the ground-state associated with $\rho$. This so 
since if $\rho$ is $v$-representable, then by the definition of the minimizer $\psi_0$, we have
\[ (\psi_0,H_0\psi_0)_{L^2} + \int_{\R^3}\rho v = e_0(v)\]
for some $v$, i.e., $\psi_0$ is the ground-state of $H(v)$. 
(A similar result holds for a minimizer of $Q(\rho,j^p)$, see Proposition \ref{vrepQ}.)

Now, let $N=1$. Note the following: $(\psi,H_0\psi)_{L^2} = \int_{\R^3} |\nabla \psi|^2dx\geq \int_{\R^3} |\nabla |\psi||^2dx$. Thus, 
for $F_{LL}(\rho)$, it is enough to minimize over the non-negative functions of $W_1$, i.e., 
$$F_{LL}(\rho)  =  \inf\Big\{ \int_{\R^3} |\nabla \psi|^2dx | \psi\in W_1, \psi\geq 0, \psi^2=\rho \Big\}.$$ We now give criteria when $\psi_0$ 
in Theorem \ref{LiebThm1} is an eigenfunction of some $H(v)$.

\begin{proposition}
(i) Let $N=1$ and $\rho\in I_1$ be such that $\psi_0$ fulfills $\Delta\psi_0\in L^2(\R^3)$ 
and $\psi_0\neq 0$ almost everywhere (a.e.), where $\psi_0\geq 0$ minimizes $\int_{\R^3}|\nabla\psi|^2$ subject to the constraint $\psi^2=\rho$. 
Then there exists a $\phi_0\in L^2(\R^3)$ and a constant $e$ such that, with $v-e = \phi_0/\rho^{1/2}$, 
$\psi_0$ satisfies
\begin{align*}
-\Delta\psi_0  + v\psi_0 = e\psi_0,
\end{align*}
and where $\int_{\R^3} v|\psi_0|^2 >-\infty$.

\noindent (ii) For $N=1$, there exists $\rho_0\in I_1$ such that $\Delta\psi_0\notin L^2(\R^3)$, and $-\Delta\psi_0  + v\psi_0 = 0$ implies 
$\int_{\R^3} v|\psi_0|^2 =-\infty$.
\label{Prop1}
\end{proposition}
\noindent{\it Proof.} By assumption, $\psi_0>0$ a.e. and $\psi_0 = \rho^{1/2}$. Now, set $\phi_0= \Delta\psi_0$, which is in $L^2(\R^3)$. Then with 
$v-e = \phi_0/\rho^{1/2}$ the conclusion of the first part follows, since 
$$\int_{\R^3}v|\psi_0|^2 = \int_{\R^3}\phi_0\rho^{1/2} + e\geq -||\phi_0 ||_{L^2} + e.$$

For the second part, set, for small $|x_1|$, $\rho_0(x)= \rho_1(x_1)\tilde{\rho}(x_2,x_3)$, where $\tilde{\rho}(x_2,x_3)$ is regular and 
$\rho_1(x_1)= (a +b|x_1|^{\varepsilon + 1/2})^2$, $a,b<0$ and $0<\varepsilon<1/2$. Then $\Delta \psi_0 \notin L^2(\R^3)$. Furthermore, 
$-\Delta\psi_0  + v\psi_0 =0$ implies $\int_{\R^3}v|\psi_0|^2=-\infty$. (The density $\rho_0$ is the counterexample of Englisch and Englisch 
that shows that not every $N$-representable density is $v$-representable, see \cite{Englisch}.)$\,\,\,\blacksquare$\\

Note that $\psi_0$ is not proven to be the ground-state of $-\Delta  + v$. However, we have

\begin{corollary}
Let $\rho$, $\psi_0$ and $\phi_0$ be as in Proposition \ref{Prop1} (i). In addition, assume that $\phi_0\leq C \rho^{1/2}$ for some constant $C$ 
and that $\rho^{-1}\in L_{\text{loc}}^1(\R^3)$. Then $\psi_0$ is the ground-state of $-\Delta + v$.
\end{corollary}
{\it Proof.} From Proposition \ref{Prop1}, we know that $-\Delta\psi_0 + v\psi_0 = e\psi_0$, where $v = \phi_0 /\rho^{1/2} +e$. 
By Schwarz's inequality, 
it follows that $v\in L_{\text{loc}}^1(\R^3)$. Since $v$ is also bounded above, we have by Corollary 11.9 
in \cite{LiebLoss} that $\psi_0>0$ is the ground-state of $-\Delta + v$. $\,\,\,\blacksquare$\\

We can thus conclude with the following characterization: 
if $\rho\in I_1$ satisfies (i) $\rho>0$ (a.e.), (ii) $\Delta\rho^{1/2}\in L^2(\R^3)$ and bounded above by a constant times $\rho^{1/2}$, 
and (iii) $\rho^{-1}\in L_{\text{loc}}^1$, then $\rho$ is $v$-representable.

\section{IV. RIGOROUS KOHN-SHAM THEORY FOR CDFT}
By means of the Levy-Lieb-type density functional $Q(\rho,j^p)$ we can formulate a rigorous $N$-representable 
Kohn-Sham approach for CDFT as that of Ref. 
\cite{Hadjisavvas84} for DFT. Now, fix the particle number $N$. We say that a wavefunction $\phi\in W_N$ is a determinant if there exist 
$N$ orthonormal one-particle functions $f_k$ such that 
\[ \phi(x_1,\dots,x_N) = (N!)^{-1/2} \det [f^k(x_l)]_{k,l}. \]
Let the space of all normalized determinants of finite kinetic energy be denoted $W_S$, i.e.,
\begin{align*}
W_S = \{\phi|\text{$\phi$ is a determinant}, ||\phi||_{L^2(\R^{3N})}=1, (\phi,K\phi)_{L^2(\R^{3N})}<\infty \},
\end{align*}
where $K= -\sum_{k=1}^N \Delta_k$. Note that, in particular, for a $\phi\in W_S$, we have $\rho_\phi = \sum_{k=1}^N |f^k|^2$ and 
\[  (\phi,K\phi)_{L^2(\R^{3N})} = \sum_{k=1}^N \int_{\R^3} |\nabla f^k|^2 dx.\]
Thus, $||\phi||_{L^2(\R^{3N})}=1$ and $(\phi,K\phi)_{L^2(\R^{3N})} <\infty$ are equivalent to $f^k\in \cH^1(\R^3)$ for all $k$.
Also note that a $\psi\in W_N$ is not in general an element of $W_S$, i.e., $W_S\subsetneq W_N$.

Furthermore, define, for a non-interacting system, the non-interacting Hamiltonian 
\begin{align*}
H'(v,A) = \sum_{k=1}^N \left((i\nabla_k - A(x_k))^2 + v(x_k) \right).
\end{align*}
The non-interacting ground-state energy is then given by
\begin{align*}
e_0'(v,A) = \inf\{\mathcal{E}_{v,A}'(\psi) | \psi\in W_N\},
\end{align*}
where $\mathcal{E}_{v,A}'(\psi)$ is given by the relation 
\[\mathcal{E}_{v,A}'(\psi) + \sum_{1\leq k<l\leq N} \int_{\R^{3N}} |\psi|^2|x_k - x_l|^{-1} =  \mathcal{E}_{v,A}(\psi).\] 
This motivates: set, for $(\rho,j^p)\in Y_N$,
\begin{align*}
Q'(\rho,j^p) = \inf\{ (\psi,K \psi)_{L^2}|\psi\in W_N,\psi\mapsto (\rho,j^p)\}.
\end{align*}

For $Q(\rho,j^p)$ and $Q'(\rho,j^p)$ we have the following.
\begin{theorem}
Fix $(\rho,j^p)\in Y_N$, then (i) there exists a $\psi_m\in W_N$ such that $\psi_m\mapsto (\rho,j^p)$ and 
$Q(\rho,j^p) = (\psi_m,H_0\psi_m)_{L^2}$, and 
(ii) there exists a $\psi_m'\in W_N$ such that $\psi_m'\mapsto (\rho,j^p)$ and 
$Q'(\rho,j^p) = (\psi_m',K\psi_m')_{L^2}$.
\label{ThmQmin}
\end{theorem}
{\it Proof.} Part (i) above is just Theorem 5 in \cite{Andre}. However, for (ii), we can use the same proof. 
For the sake of completeness we include the proof in \cite{Andre} here applied to $Q'(\rho,j^p)$.

Let $\{ \psi^j \}_{j=1}^\infty$ be a minimizing sequence, i.e., 
$\psi^j\in W_N$, $\psi^j\mapsto (\rho,j^p)$ and $$\lim_{j\rightarrow \infty} (\psi^j,K\psi^j)_{L^2}= Q'(\rho,j^p).$$
Since $\{ \psi^j \}_{j=1}^\infty$ is bounded in $\cH^1(\R^{3N})$, by the Banach-Alaoglu theorem there exists a 
subsequence and a $\psi_m'\in \cH^1(\R^{3N})$ such that $\psi^{j_k}\rightharpoonup \psi_m'$ weakly in $\cH^1(\R^{3N})$ as $k\rightarrow \infty$. 
Since the functional $\psi\mapsto(\psi,K\psi)_{L^2}$ is weakly lower semi continuous, we know that 
\[(\psi_m',K\psi_m')_{L^2}\leq Q'(\rho,j^p).\] 
However, it remains to prove that 
$\psi_m'\mapsto (\rho,j^p)$. In the proof of Theorem 3.3 in \cite{Lieb83}, it is shown that 
$\psi^{j_k}\rightarrow \psi_m'$ in $L^2(\R^{3N})$ and $\psi_m'\mapsto \rho$. Now, let $g$ be the 
characteristic function of any measurable set in $\R^3$.  For $l=1,2,3$ and $k=1,2,\dots$, let
\[ I_{l}(k) = \left| \int_{\R^{3N}} [(\psi^{j_k})^* \partial_l\psi^{j_k}  -(\psi_m')^*\partial_l\psi_m'] g\right|.\]
Then
\begin{align*}
I_l(k) &\leq 
\left|\int_{\R^{3N}} (\psi^{j_k} - \psi_m')^* (\partial_l\psi^{j_k})g\right|
+ \left| \int_{\R^{3N}} (\psi_m')^* (\partial_l\psi^{j_k} - \partial_l \psi_m')g \right| \\
&\leq ||  \psi^{j_k} - \psi_m'  ||_{L^2} ||  (\partial_l\psi^{j_k})g ||_{L^2} + 
\left| \int_{\R^{3N}} (\psi_m' g^*)^* (\partial_l\psi^{j_k} - \partial_l \psi_m') \right|.
\end{align*}
Thus $I_l(k)$ tends to zero as $k\rightarrow \infty$ (because $\psi^{j_k}\rightarrow \psi_m'$ in 
$L^2(\R^{3N})$-norm and $\psi^{j_k}\rightharpoonup \psi_m'$ weakly in $\cH^1(\R^{3N})$ as $k\rightarrow \infty$). Since 
$\psi^{j_k}\mapsto j^p$ for all $k$, we 
have $\int_{\R^3} (j^p)_l g = \int_{\R^3} (j_{\psi_m'}^p)_l g$, i.e., $ j_{\psi_m'}^p(x) = j^p(x)$ a.e. $\,\,\,\blacksquare$

\begin{proposition}
Assume that $(\rho,j^p)\in \cA_N$, i.e., there exists a $H(v,A)$ with ground-state $\psi$ such that $\psi\mapsto (\rho,j^p)$. 
Then the minimizer $\psi_m$ is the ground-state of $H(v,A)$.   
\label{vrepQ}
\end{proposition}
{\it Proof.} Since $\psi\mapsto (\rho,j^p)$, we have $(\psi,H_0\psi)_{L^2} \geq (\psi_m,H_0\psi_m)_{L^2}$. The 
conclusion then follows from
\begin{align*}
e_0(v,A) &\leq (\psi_m,H(v,A)\psi_m)_{L^2} = (\psi_m,H_0\psi_m)_{L^2} + 2\int_{\R^3}j^p\cdot A + \int_{\R^3}\rho(v + |A|^2) \\
&\leq (\psi,H_0\psi)_{L^2} + 2\int_{\R^3}j^p\cdot A + \int_{\R^3}\rho(v + |A|^2) = (\psi,H(v,A)\psi)_{L^2} = e_0(v,A).\,\,\,\blacksquare
\end{align*}

Note that when $H_0$ is replaced by $K$, $Q'(\rho,j^p)$ is the minimal kinetic energy for $\psi\in W_N$ such 
that $\rho_\psi = \rho $ and $j_\psi^p = j^p$. Next we will introduce another kinetic energy density functional.

\subsection{A. Non-interacting kinetic energy density functional}
Set, for $(\rho,j^p)\in Y_N$,
\begin{align*}
T_{\text{det}}(\rho,j^p) = \inf\{ (\phi,K \phi)_{L^2}| \phi\in W_S,\phi\mapsto (\rho,j^p)\}.
\end{align*}
For $(\rho,j^p)\in Y_N$, we remark that the set $\{ \phi\in W_S| \phi\mapsto (\rho,j^p)\}$ is not empty, at least when $N\geq 4$. This 
follows from the determinant construction in \cite{LiebSchrader}. However, for all $N$, the set 
$\{ \phi\in W_S| \phi\mapsto (\rho,j^p),\nabla \times (j^p/\rho) =0\}$ is non-empty (see \cite{LiebSchrader,Andre}).
%\begin{theorem}
%For $(\rho,j^p)\in Y_N$, there exists a $\psi_m'\in W_N$ such that $\psi_m'\mapsto (\rho,j^p)$ and 
%$Q'(\rho,j^p) = (\psi_m',K\psi_m')_{L^2}$.
%\label{ThmQprim}
%\end{theorem}

We have that $T_{\text{det}}(\rho,j^p)\geq Q'(\rho,j^p)$ on $Y_N$. Now, let the set of non-interacting $v$-representable densities be denoted $\cA_N'$,
\begin{align*}
\cA_N' = \{(\rho,j^p)| H'(v,A) \text{ has a unique ground-state}\}.
\end{align*}
If $(\rho,j^p)\in \cA_N'$, by the same argument as in the proof of Proposition \ref{vrepQ}, we can conclude that $\psi_m'$ is 
the ground-state of some $H'(v,A)$. Clearly, $\psi_m'$ is in this case a determinant. Thus, 
$T_{\text{det}}(\rho,j^p) = Q'(\rho,j^p)$ on $\cA_N'$. 
%(The minimizer $\psi_m'$ of $Q'$ is in that case a determinant and it is also 
%the non-degenerate ground-state of some $H'(v,A)$ having ground-state densities $\rho$ and $j^p$.)

An important property of $T_{\text{det}}(\rho,j^p)$ is that the infimum actually is a minimum. For the proof, we need the following:

(i) For $k=1,\dots,N$, assume that $f_j^k \rightarrow f^k$ in $L^2$-norm as $j\rightarrow \infty$ and for each $j$, $(f_j^k,f_j^l)_{L^2}= \delta_{kl}$. Then
$f^1,\dots,f^N$ are orthonormal. This so since
\begin{align*}
(f^k,f^l)_{L^2} = \lim_{j\rightarrow \infty}(f_j^k,f^l)_{L^2}=
\lim_{j\rightarrow \infty}[(f_j^k,f^l -f_j^l)_{L^2} + (f_j^k,f_j^l)_{L^2}] = \delta_{kl},
\end{align*}
where we used that $|(f_j^k,f^l -f_j^l)_{L^2}| \leq || f_j^k ||_{L^2} ||f^l -f_j^l ||_{L^2}\rightarrow 0$ as $j\rightarrow \infty$.

(ii) If $f_j \rightharpoonup f$ weakly in $L^2$ as $j\rightarrow \infty$ and $||f_j||_{L^2} \rightarrow ||f||_{L^2}$ as $j\rightarrow \infty$, 
then $f_j \rightarrow f$ in $L^2$-norm as $j\rightarrow \infty$. (This is an elementary fact and can be checked by expanding 
$||f_j-f||_{L^2}^2 =(f_j-f,f_j-f)_{L^2}$.)

\begin{theorem}
Let $(\rho,j^p)\in Y_N$. If $N<4$ we also assume $\nabla \times (j^p/\rho) =0$. Then there 
exists a determinant $\phi_m$ such that $\phi_m\mapsto (\rho,j^p)$ and 
$T_{\text{det}}(\rho,j^p) = (\phi_m,K\phi_m)_{L^2}$.
\label{ThmTdet}
\end{theorem}
{\it Proof.} Fix $(\rho,j^p)\in Y_N$ and let $\{D^j\}_{j=1}^\infty\subset W_S$ be a sequence of minimizing determinants, i.e., $D^j\mapsto (\rho,j^p)$ and 
$\lim_{j\rightarrow \infty}(D^j,K D^j)_{L^2} =T_{\text{det}}(\rho,j^p)$. From the proof of Theorem \ref{ThmQmin}, there exists a subsequence 
$D^{j_n}$ and a $\phi_m\in W_N$ such that 
$\phi_m\mapsto (\rho,j^p)$, 
\[T_{\text{det}}(\rho,j^p) = (\phi_m,K \phi_m)_{L^2} \] 
and $D^{j_n}\rightarrow \phi_m$ in $L^2$-norm. It remains to show that $\phi_m\in W_S$. To meet that end, let
\[D^j(x_1,\dots,x_N) = (N!)^{-1/2}\det[f_j^k(x_l)]_{k,l}, \]
where for each $j$ the $N$ one-particle functions $f_j^k$ are orthonormal. By the Banach-Alaoglu theorem, there 
exist $N$ functions $f^k$ such that (for a subsequence) $f_j^k \rightharpoonup f^k$ weakly in $L^2$ as $j\rightarrow \infty$. 
We furthermore claim that $f^1,\dots,f^N$ are orthonormal. If we could prove 
that $f_j^k \rightarrow f^k$ in $L^2$-norm, it would follow that $(f^k,f^l)_{L^2}= \delta_{kl}$. 

We shall prove $f_j^k \rightarrow f^k$ by demonstrating that $||f_j^k||_{L^2} \rightarrow ||f^k||_{L^2}$. 
This together with the fact that $f_j^k \rightharpoonup f^k$ weakly in $L^2$ gives the desired result. Let $\eps>0$ and 
choose a characteristic function $\chi$ such that $\int_{\R^3}\rho (1-\chi)<\eps $. Since for each $j$, 
$D^j \mapsto \rho$, we have for each $k$,
\[  \int_{\R^3} |f_j^k|^2 (1-\chi) \leq \sum_{k=1}^N \int_{\R^3} |f_j^k|^2 (1-\chi) =\int_{\R^3}\rho (1-\chi)<\eps.\]
By the Rellich-Kondrachov theorem, we can choose a subsequence such that 
$\chi f_{j_n}^k\rightarrow \chi f^k$ in $L^2$-norm. But this implies
\[  \int_{\R^3} |f^k|^2 \geq  \int_{\R^3} \chi|f^k|^2
= \lim_{n\rightarrow \infty} \int_{\R^3} \chi|f_{j_n}^k|^2\geq 1-\eps. \]
Conversely, by the lower semi continuity of the $L^2$-norm, $1=\lim \inf_{j\rightarrow\infty} ||f_j^k||_{L^2}\geq ||f^k||_{L^2}$, and we have 
$||f^k||_{L^2}=1$.

Returning to the fact that $f_{j_n}^k \rightharpoonup f^k$ weakly in $L^2$, we note that 
$\Pi_{k=1}^N f_{j_n}^k (x_k)\rightharpoonup \Pi_{k=1}^N f^k (x_k)$ weakly in $L^2(\R^{3N})$ (since product-functions are dense in $L^2(\R^{3N})$). 
But then 
\[D^{j_n}\rightharpoonup (N!)^{-1/2}\det[f^k(x_l)]_{k,l},\] 
where $f^1,\dots,f^N$ are orthonormal. However, since $D^{j_n}\rightarrow \phi_m$, we have $\phi_m \in W_S$. $\,\,\,\blacksquare$\\

\subsection{B. $N$-representable Kohn-Sham theory}
In the Kohn-Sham approach \cite{KS65}, a non-interacting system is introduced that has the 
same ground-state density as the fully interacting system. 
The idea is then to use an element of $W_S$, i.e., a determinant, to compute the ground-state density. 
On $\cA_N'$, the (generalized) Kohn-Sham density functional $T_{KS}(\rho,j^p)$ satisfies 
\[T_{KS}(\rho,j^p)= T_{\text{det}}(\rho,j^p)= Q'(\rho,j^p).\] 
Moreover, $T_{KS}$ defines an exchange-correlation functional $E_{xc}(\rho,j^p)$ on $\cA_N\cap \cA_N'$ 
according to
\begin{align*}
E_{xc}(\rho,j^p) = F_{HK}(\rho,j^p) - \frac{1}{2}\int_{\R^3}\int_{\R^3}\frac{\rho(x)\rho(y)}{|x-y|}dxdy - T_{KS}(\rho,j^p).
\end{align*}

Now, to obtain an $N$-representable Kohn-Sham scheme, define two functionals on $W_S$,
\begin{align*}
\G_K(\phi) &= \inf\{ (f,Kf)_{L^2}| f\in W_S, f\mapsto (\rho_\phi,j_\phi^p)\}, \\
\G_{H_0}(\phi) &=\inf\{ (f,H_0f)_{L^2}| f\in W_N, f\mapsto (\rho_\phi,j_\phi^p)\}.
\end{align*}
Note that, by Theorem \ref{ThmQmin} and Theorem \ref{ThmTdet}, there exists a $\psi_m\in W_N$ and a $\phi_m\in W_S$ such that 
$\G_{H_0}(\phi)= (\psi_m,H_0\psi_m)_{L^2}$ and $\G_K(\phi)= (\phi_m,K\phi_m)_{L^2}$ 
and where $\psi_m,\phi_m \mapsto (\rho_\phi,j_\phi^p)$. Furthermore, we can use the existence of the minimizers $\psi_m$ and 
$\phi_m$ and define, for $\phi\in W_S$, 
\begin{align*}
\Delta T(\phi) &= (\psi_m,K\psi_m)_{L^2} - (\phi_m,K\phi_m)_{L^2},\\
E_{xc}^W(\phi) &= (\psi_m, \sum_{1\leq k<l\leq N}|x_k-x_l|^{-1}\psi_m)_{L^2} - \frac{1}{2}\int_{\R^3}\int_{\R^3}\frac{\rho_\phi(x)\rho_\phi(y)}{|x-y|}dxdy.
\end{align*}
On $W_S$, we now introduce the following energy functional
\begin{align*}
\G_{v,A}(\phi)&= (\phi,K\phi)_{L^2} + \Delta T(\phi)+2\int_{\R^3}j_{\phi}^p\cdot A \\
&+ \int_{\R^3}\rho_{\phi}(v + |A|^2)  + E_{xc}^W(\phi) + \frac{1}{2}\int_{\R^3}\int_{\R^3}\frac{\rho_\phi(x)\rho_\phi(y)}{|x-y|}dxdy.
\end{align*}
We then have 

\begin{theorem}
Assume that $H(v,A)$ has a unique ground-state $\psi_0$. Let $e_0(v,A)$, $\rho_0$ and $j_0^p$ denote the ground-state energy, ground-state 
particle density and ground-state paramagnetic current density, respectively. If $N<4$ we assume that $\nabla \times (j_0^p/\rho_0)=0$. Then 
\[ e_0(v,A) = \inf\{ \G_{v,A}(\phi)| \phi\in W_S \} =  \G_{v,A}(\phi_m)\]
for some $\phi_m\in W_S$. Moreover, $\rho_{\phi_m} = \rho_0$ and $j_{\phi_m}^p= j_0^p$, i.e., the ground-state densities can be computed from 
the determinant $\phi_m$ that minimizes $\G_{v,A}$.
\label{NrepThm}
\end{theorem}
{\it Proof.} First note, for any $\phi\in W_S$, we have
\begin{align*}
\G_{v,A}(\phi) &= (\phi,K\phi)_{L^2} + \left((\psi_m,K\psi_m)_{L^2} - (\phi_m,K\phi_m)_{L^2}\right)
\\&+2\int_{\R^3}j_\phi^p\cdot A+\int_{\R^3}\rho_\phi(v + |A|^2) +(\psi_m,\sum_{1\leq k<l\leq N}|x_k-x_l|^{-1}\psi_m)_{L^2}\\
&\geq  (\psi_m,(K+\sum_{1\leq k<l\leq N}|x_k-x_l|^{-1})\psi_m)_{L^2}+2\int_{\R^3}j_\phi^p\cdot A+\int_{\R^3}\rho_\phi(v + |A|^2) \\
&=\mathcal{E}_{v,A}(\psi_m)\geq e_0(v,A),
\end{align*}
where we used that $(\phi,K\phi)_{L^2}- (\phi_m,K\phi_m)_{L^2}\geq 0$ and $\psi_m\mapsto (\rho_\phi,j_\phi^p)$. In the next step, 
we want to show that there exists a $\phi_0\in W_S$ such that $\G_{v,A}(\phi_0)=e_0(v,A)$ and $\phi_0\mapsto (\rho_0,j_0^p)$.

Let $\phi\in W_S$ be a determinant such that $\phi\mapsto (\rho_0,j_0^p)$ (if $N<4$, we need the assumption $\nabla \times (j_0^p/\rho_0)=0$). 
By Theorem \ref{ThmTdet}, we then have
\begin{align*}
\G_K(\phi) = T_{\text{det}}(\rho_0,j_0^p) = (\phi_m,K\phi_m)_{L^2},
\end{align*}
for some $\phi_m\in W_S$. Note that $\phi_m$ is a determinant such that $\phi_m\mapsto (\rho_0,j_0^p)$ and 
\[\G_K(\phi_m)= (\phi_{m,m},K\phi_{m,m})_{L^2}= (\phi_m,K\phi_m)_{L^2}.\] 
Furthermore, 
\begin{align*}
\G_{H_0}(\phi_m) = Q(\rho_0,j_0^p) = (\psi_m,H_0\psi_m)_{L^2},
\end{align*}
for some $\psi_m\in W_N$, which follows from Theorem \ref{ThmQmin}. 
Note that $\psi_m \mapsto  (\rho_0,j_0^p)=(\rho_{\phi_m},j_{\phi_m}^p)$. We have, 
\begin{align*}
e_0(v,A) &= (\psi_m,H(v,A)\psi_m)_{L^2} \\
&= (\psi_m,H_0\psi_m)_{L^2}+2\int_{\R^3}j_0^p\cdot A+\int_{\R^3}\rho_0(v + |A|^2) \\ 
&= (\psi_m,K\psi_m)_{L^2} +(\psi_m, \sum_{1\leq k<l \leq N}|x_k-x_l|^{-1}\psi_m)_{L^2}
+2\int_{\R^3}j_{\phi_m}^p\cdot A+\int_{\R^3}\rho_{\phi_m}(v + |A|^2),
\end{align*}
where the first equality follows from Proposition \ref{vrepQ}. 
Since \[\Delta T(\phi_m) = (\psi_m,K\psi_m)_{L^2} - (\phi_m,K\phi_m)_{L^2}\] and 
\[ E_{xc}^W(\phi_m) = (\psi_m, \sum_{1\leq k<l\leq N}|x_k-x_l|^{-1}\psi_m)_{L^2} - 
\frac{1}{2}\int_{\R^3}\int_{\R^3}\frac{\rho_{\phi_m}(x)\rho_{\phi_m}(y)}{|x-y|}dxdy,
\]
it follows that
\begin{align*}
e_0(v,A)&= (\phi_m,K\phi_m)_{L^2} +2\int_{\R^3}j_{\phi_m}^p\cdot A + \int_{\R^3}\rho_{\phi_m}(v + |A|^2)
  \\& + \frac{1}{2}\int_{\R^3}\int_{\R^3}\frac{\rho_{\phi_m}(x)\rho_{\phi_m}(y)}{|x-y|}dxdy+ E_{xc}^W(\phi_m)+ \Delta T(\phi_m) = \G_{v,A}(\phi_m).
  \,\,\,\blacksquare
\end{align*}
\\
\noindent \textbf{Remarks.} (i) Any density pair $(\rho,j^p)$ computed from a $\phi\in W_S$ is $N$-representable, but not necessarily 
(non-interacting) $v$-representable. So Theorem \ref{NrepThm} establishes a Kohn-Sham approach for $N$-representable densities 
(whereas $T_{KS}$ is only defined on $\cA_N'$).

(ii) Recall that no Hohenberg-Kohn theorem can exist for CDFT formulated with the 
paramagnetic current density. On the other hand, since $\rho$ and $j^p$ determine the ground-state, 
the Hohenberg-Kohn variational principle continues to hold for CDFT formulated with these densities. 
However, the $N$-representable Kohn-Sham approach outlined here does not use any variational principle for densities. Instead, the 
approach relies on the existence of minimizers for certain functionals.

(iii) If we set $\phi(x_1,\dots,x_N) = (N!)^{-1/2}\det[f^k(x_l) ]_{k,l}$ and 
define on $(\cH^1(\R^3))^N$ the functional
\begin{align*}
\mathcal{E}(f^1,\dots,f^N) &= \sum_{k=1}^N \int_{\R^3}|\nabla f^k|^2  + 2\sum_{k=1}^N\int_{\R^3}
\text{Im}(\overline{f}^k\nabla f^k)\cdot A + \sum_{k=1}^N \int_{\R^3}|f^k|^2(v+|A|^2)\\
&+\frac{1}{2}\sum_{k,l=1}^N \int_{\R^3}\int_{\R^3}\frac{|f^k(x)|^2|f^l(y)|^2}{|x-y|}dxdy + E_{xc},
\end{align*}
where $E_{xc} = \Delta T + E_{xc}^W$, we can obtain the usual Kohn-Sham equations by 
minimizing $\mathcal{E}(f^1,\dots,f^N)$ subject to the constraint $(f^k,f^l)_{L^2}= \delta_{kl}$.

\section{ACKNOWLEDGMENTS}
The author is thankful to Michael Benedicks for useful comments and discussions.

\bibliography{./KSreferences}

%Merlin.mbs v4.21 2009-07-09.
\begin{thebibliography}{10}%
\makeatletter
\providecommand \@ifxundefined [1]{%
 \ifx #1\undefined \expandafter \@firstoftwo
 \else \expandafter \@secondoftwo
\fi
}%
\providecommand \@ifnum [1]{%
 \ifnum #1\expandafter \@firstoftwo
 \else \expandafter \@secondoftwo
\fi
}%
\providecommand \enquote [1]{``#1''}%
\providecommand \bibnamefont  [1]{#1}%
\providecommand \bibfnamefont [1]{#1}%
\providecommand \citenamefont [1]{#1}%
\providecommand\href[0]{\@sanitize\@href}%
\providecommand\@href[1]{\endgroup\@@startlink{#1}\endgroup\@@href}%
\providecommand\@@href[1]{#1\@@endlink}%
\providecommand \@sanitize [0]{\begingroup\catcode`\&12\catcode`\#12\relax}%
\@ifxundefined \pdfoutput {\@firstoftwo}{%
 \@ifnum{\z@=\pdfoutput}{\@firstoftwo}{\@secondoftwo}%
}{%
 \providecommand\@@startlink[1]{\leavevmode\special{html:<a href="#1">}}%
 \providecommand\@@endlink[0]{\special{html:</a>}}%
}{%
 \providecommand\@@startlink[1]{%
  \leavevmode
  \pdfstartlink
   attr{/Border[0 0 1 ]/H/I/C[0 1 1]}%
   user{/Subtype/Link/A<</Type/Action/S/URI/URI(#1)>>}%
  \relax
 }%
 \providecommand\@@endlink[0]{\pdfendlink}%
}%
\providecommand \url  [0]{\begingroup\@sanitize \@url }%
\providecommand \@url [1]{\endgroup\@href {#1}{\urlprefix}}%
\providecommand \urlprefix [0]{URL }%
\providecommand \Eprint[0]{\href }%
\@ifxundefined \urlstyle {%
  \providecommand \doi [1]{doi:\discretionary{}{}{}#1}%
}{%
  \providecommand \doi [0]{doi:\discretionary{}{}{}\begingroup
  \urlstyle{rm}\Url }%
}%
\providecommand \doibase [0]{http://dx.doi.org/}%
\providecommand \Doi[1]{\href{\doibase#1}}%
\providecommand \bibAnnote [3]{%
  \BibitemShut{#1}%
  \begin{quotation}\noindent
    \textsc{Key:}\ #2\\\textsc{Annotation:}\ #3%
  \end{quotation}%
}%
\providecommand \bibAnnoteFile [2]{%
  \IfFileExists{#2}{\bibAnnote {#1} {#2} {\input{#2}}}{}%
}%
\providecommand \typeout [0]{\immediate \write \m@ne }%
\providecommand \selectlanguage [0]{\@gobble}%
\providecommand \bibinfo [0]{\@secondoftwo}%
\providecommand \bibfield [0]{\@secondoftwo}%
\providecommand \translation [1]{[#1]}%
\providecommand \BibitemOpen[0]{}%
\providecommand \bibitemStop [0]{}%
\providecommand \bibitemNoStop [0]{.\EOS\space}%
\providecommand \EOS [0]{\spacefactor3000\relax}%
\providecommand \BibitemShut [1]{\csname bibitem#1\endcsname}%
%</preamble>
\bibitem{HK64}%
  \BibitemOpen
  \bibfield{author}{%
  \bibinfo {author} {\bibfnamefont{P.}~\bibnamefont{Hohenberg}}\ and\ \bibinfo
  {author} {\bibfnamefont{W.}~\bibnamefont{Kohn}},\ }%
  \bibfield{journal}{%
  \Doi{10.1103/PhysRev.136.B864}{\bibinfo {journal} {Phys. Rev.}}\ }%
  \textbf{\bibinfo {volume} {136}},\ \bibinfo {pages} {B864} (\bibinfo {month}
  {Nov}\ \bibinfo {year} {1964}),\
  \url{http://link.aps.org/doi/10.1103/PhysRev.136.B864}%
  \bibAnnoteFile{NoStop}{HK64}%
\bibitem{KS65}%
  \BibitemOpen
  \bibfield{author}{%
  \bibinfo {author} {\bibfnamefont{W.}~\bibnamefont{Kohn}}\ and\ \bibinfo
  {author} {\bibfnamefont{L.~J.}\ \bibnamefont{Sham}},\ }%
  \bibfield{journal}{%
  \Doi{10.1103/PhysRev.140.A1133}{\bibinfo {journal} {Phys. Rev.}}\ }%
  \textbf{\bibinfo {volume} {140}},\ \bibinfo {pages} {A1133} (\bibinfo {month}
  {Nov}\ \bibinfo {year} {1965}),\
  \url{http://link.aps.org/doi/10.1103/PhysRev.140.A1133}%
  \bibAnnoteFile{NoStop}{KS65}%
\bibitem{Hadjisavvas84}%
  \BibitemOpen
  \bibfield{author}{%
  \bibinfo {author} {\bibfnamefont{N.}~\bibnamefont{Hadjisavvas}}\ and\
  \bibinfo {author} {\bibfnamefont{A.}~\bibnamefont{Theophilou}},\ }%
  \bibfield{journal}{%
  \Doi{10.1103/PhysRevA.30.2183}{\bibinfo {journal} {Phys. Rev. A}}\ }%
  \textbf{\bibinfo {volume} {30}},\ \bibinfo {pages} {2183} (\bibinfo {month}
  {Nov}\ \bibinfo {year} {1984}),\
  \url{http://link.aps.org/doi/10.1103/PhysRevA.30.2183}%
  \bibAnnoteFile{NoStop}{Hadjisavvas84}%
\bibitem{Vignale2002}%
  \BibitemOpen
  \bibfield{author}{%
  \bibinfo {author} {\bibfnamefont{K.}~\bibnamefont{Capelle}}\ and\ \bibinfo
  {author} {\bibfnamefont{G.}~\bibnamefont{Vignale}},\ }%
  \bibfield{journal}{%
  \Doi{10.1103/PhysRevB.65.113106}{\bibinfo {journal} {Phys. Rev. B}}\ }%
  \textbf{\bibinfo {volume} {65}},\ \bibinfo {pages} {113106} (\bibinfo {month}
  {Feb}\ \bibinfo {year} {2002}),\
  \url{http://link.aps.org/doi/10.1103/PhysRevB.65.113106}%
  \bibAnnoteFile{NoStop}{Vignale2002}%
\bibitem{AndreMichael}%
  \BibitemOpen
  \bibfield{author}{%
  \bibinfo {author} {\bibfnamefont{A.}~\bibnamefont{Laestadius}}\ and\ \bibinfo
  {author} {\bibfnamefont{M.}~\bibnamefont{Benedicks}},\ }%
  \bibinfo {journal} {To appear in Int. Jour. Quant. Chem.}%
  \bibAnnoteFile{Stop}{AndreMichael}%
\bibitem{Tellgren}%
  \BibitemOpen
\bibfield{journal}{%
    }%
  \bibfield{author}{%
  \bibinfo {author} {\bibfnamefont{E.~I.}\ \bibnamefont{Tellgren}}, \bibinfo
  {author} {\bibfnamefont{S.}~\bibnamefont{Kvaal}}, \bibinfo {author}
  {\bibfnamefont{E.}~\bibnamefont{Sagvolden}}, \bibinfo {author}
  {\bibfnamefont{U.}~\bibnamefont{Ekstr\"om}}, \bibinfo {author}
  {\bibfnamefont{A.~M.}\ \bibnamefont{Teale}},\ and\ \bibinfo {author}
  {\bibfnamefont{T.}~\bibnamefont{Helgaker}},\ }%
  \bibfield{journal}{%
  \Doi{10.1103/PhysRevA.86.062506}{\bibinfo {journal} {Phys. Rev. A}}\ }%
  \textbf{\bibinfo {volume} {86}},\ \bibinfo {pages} {062506} (\bibinfo {month}
  {Dec}\ \bibinfo {year} {2012}),\
  \url{http://link.aps.org/doi/10.1103/PhysRevA.86.062506}%
  \bibAnnoteFile{NoStop}{Tellgren}%
\bibitem{Vignale87}%
  \BibitemOpen
  \bibfield{author}{%
  \bibinfo {author} {\bibfnamefont{G.}~\bibnamefont{Vignale}}\ and\ \bibinfo
  {author} {\bibfnamefont{M.}~\bibnamefont{Rasolt}},\ }%
  \bibfield{journal}{%
  \Doi{10.1103/PhysRevLett.59.2360}{\bibinfo {journal} {Phys. Rev. Lett.}}\ }%
  \textbf{\bibinfo {volume} {59}},\ \bibinfo {pages} {2360} (\bibinfo {month}
  {Nov}\ \bibinfo {year} {1987}),\
  \url{http://link.aps.org/doi/10.1103/PhysRevLett.59.2360}%
  \bibAnnoteFile{NoStop}{Vignale87}%
\bibitem{Lieb83}%
  \BibitemOpen
  \bibfield{author}{%
  \bibinfo {author} {\bibfnamefont{E.~H.}\ \bibnamefont{Lieb}},\ }%
  \bibfield{journal}{%
  \Doi{10.1002/qua.560240302}{\bibinfo {journal} {Int. Jour. Quant. Chem.}}\ }%
  \textbf{\bibinfo {volume} {24}},\ \bibinfo {pages} {243} (\bibinfo {year}
  {1983}),\ \url{http://dx.doi.org/10.1002/qua.560240302}%
  \bibAnnoteFile{NoStop}{Lieb83}%
\bibitem{Englisch}%
  \BibitemOpen
  \bibfield{author}{%
  \bibinfo {author} {\bibfnamefont{H.}~\bibnamefont{Englisch}}\ and\ \bibinfo
  {author} {\bibfnamefont{R.}~\bibnamefont{Englisch}},\ }%
  \bibfield{journal}{%
  \Doi{10.1016/0378-4371(83)90254-6}{\bibinfo {journal} {Physica A}}\ }%
  \textbf{\bibinfo {volume} {121}},\ \bibinfo {pages} {253} (\bibinfo {year}
  {1983})%
  \bibAnnoteFile{NoStop}{Englisch}%
\bibitem{Andre}%
  \BibitemOpen
  \bibfield{author}{%
  \bibinfo {author} {\bibfnamefont{A.}~\bibnamefont{Laestadius}},\ }%
  \bibfield{journal}{%
  \bibinfo {journal} {arXiv}\ }%
  \textbf{\bibinfo {volume} {abs/1404.0825}} (\bibinfo {year} {2014}),\
  \url{http://arxiv.org/abs/1404.0825}%
  \bibAnnoteFile{NoStop}{Andre}%
\bibitem{LiebLoss}%
  \BibitemOpen
  \bibfield{author}{%
  \bibinfo {author} {\bibfnamefont{E.~H.}\ \bibnamefont{Lieb}}\ and\ \bibinfo
  {author} {\bibfnamefont{M.}~\bibnamefont{Loss}},\ }%
  \emph{\bibinfo {title} {Analysis}}\ (\bibinfo {publisher} {American
  Mathematical Society},\ \bibinfo {address} {Providence, Rhode Island},\
  \bibinfo {year} {2001})%
  \bibAnnoteFile{NoStop}{LiebLoss}%
\bibitem{LiebSchrader}%
  \BibitemOpen
  \bibfield{author}{%
  \bibinfo {author} {\bibfnamefont{E.~H.}\ \bibnamefont{Lieb}}\ and\ \bibinfo
  {author} {\bibfnamefont{R.}~\bibnamefont{Schrader}},\ }%
  \bibfield{journal}{%
  \Doi{10.1103/PhysRevA.88.032516}{\bibinfo {journal} {Phys. Rev. A}}\ }%
  \textbf{\bibinfo {volume} {88}},\ \bibinfo {pages} {032516} (\bibinfo {month}
  {Sep}\ \bibinfo {year} {2013}),\
  \url{http://link.aps.org/doi/10.1103/PhysRevA.88.032516}%
  \bibAnnoteFile{NoStop}{LiebSchrader}%
\end{thebibliography}%

\end{document}